# A Data Structure Perspective to the RDD-based Apriori Algorithm on Spark

Pankaj Singh[1], Sudhakar Singh[2], P. K. Mishra[1], and Rakhi Garg[3]

[1] Department of Computer Science, Banaras Hindu University, Varanasi, India
{psingh.edu,mishra}@bhu.ac.in

[2] Department of Electronics and Communication, University of Allahabad, Prayagraj, India
sudhakar@allduniv.ac.in

[3] Mahila Maha Vidyalaya, Banaras Hindu University, Varanasi, India
rgarg@bhu.ac.in

**Abstract**

During the recent years, a number of efficient and scalable frequent itemset mining algorithms for big data analytics have been proposed by many researchers. Initially, MapReduce-based frequent itemset mining algorithms on Hadoop cluster were proposed. Although, Hadoop has been developed as a cluster computing system for handling and processing big data, but the performance of Hadoop does not meet the expectation for the iterative algorithms of data mining, due to its high I/O, and writing and then reading intermediate results in the disk. Consequently, Spark has been developed as another cluster computing infrastructure which is much faster than Hadoop due to its in-memory computation. It is highly suitable for iterative algorithms and supports batch, interactive, iterative, and stream processing of data. Many frequent itemset mining algorithms have been re-designed on the Spark, and most of them are Apriori-based. All these Spark-based Apriori algorithms use Hash Tree as the underlying data structure. This paper investigates the efficiency of various data structures for the Spark-based Apriori. Although, the data structure perspective has been investigated previously, but for MapReduce-based Apriori, and it must be re-investigated in the distributed computing environment of Spark. The considered underlying data structures are Hash Tree, Trie, and Hash Table Trie. The experimental results on the benchmark datasets show that the performance of Spark-based Apriori with Trie and Hash Table Trie are almost similar but both perform many times better than Hash Tree in the distributed computing environment of Spark.

## 1. Introduction

In the age of data being new oil, fast and scalable data refinement techniques are imperative. The refinement techniques for data refer to the process of discovering and interpreting hidden and meaningful patterns in the data. Association rule mining (ARM) [1] is such a kind of technique that discovers the interesting correlations in a transactional dataset. The computation intensive part of association rule mining is the frequent itemsets generation, as the association rules are generated from frequent itemsets. So, the various frequent itemset mining (FIM) algorithms have been developed to generate frequent itemsets. Three well known basic FIM algorithms are Apriori [1], Eclat [2], and FP-Growth [3]. A number of improved versions of these algorithms have been developed by the various researchers to speed up the generation of frequent itemsets. Being sequential computing nature of these algorithms, they could not be applied to process big data set. Parallel and distributed versions of FIM algorithms [4] have been developed to generate frequent itemsets for the larger datasets. Although, these algorithms are efficient, but designed for the traditional distributed system. The problem with traditional distributed system is that it wastes the processing power in data movement between computing nodes that also consumes high network bandwidth. Fault tolerance is also not supported, and failed jobs need to be restarted again. Apache Hadoop [5], a cluster computing system overcomes all these problems. Hadoop is a large scale distributed batch processing system, designed to handle big data. It is a fault tolerant system based on the share nothing architecture. It moves the computation to the nodes having data during the execution rather than data to the nodes. Data storage and computation are handled by the two core components of Hadoop named HDFS (Hadoop Distributed File System) [6] and MapReduce [7] respectively. MapReduce is a parallel programming paradigm that is extremely scalable and fault tolerant. HDFS breaks the big data in the form of blocks and stores them across the different nodes of the cluster in a replicated manner. An application is submitted as a MapReduce job which is splitted into multiple independent tasks and executed concurrently on different blocks of data across the cluster. A number of FIM and other data mining algorithms have been re-designed on the MapReduce framework to provide the scalability for big data. Apriori-based FIM algorithms have been attempted by most of the authors [8]. No doubt, Hadoop provides scalability and fault tolerance, but degrades the execution speed due to the high I/O, and writing and then reading the intermediate results in HDFS for the iterative algorithms. Data mining and machine learning algorithms are highly iterative, and that motivates the researchers to develop a new cluster computing architecture, Spark [9-10] that overcomes all these problems.

Apache Spark [9-10] is a fast and general engine for big data processing. Spark is much faster than Hadoop MapReduce, 100 times faster in memory and 10 times on disk [9]. It provides batch, interactive, iterative, and stream processing of data. It supports in-memory computation, and storage of the intermediate results in memory using the Resilient Distributed Datasets (RDDs) [11]. So, a number of data mining and machine learning algorithms have been re-designed on the Spark RDD framework. FIM algorithms on the Spark have been also proposed by many authors [12-18], where most of the efforts have been made on the efficient implementations of Apriori-based FIM algorithm on the Spark. The efficiency of the Spark-based Apriori algorithms extensively depend on the way it is parallelized on the Spark, and the underlying data structures used to store and compute frequent itemsets. Most of the Spark-based Apriori algorithms still use Hash Tree data structure, and did not focus on the other more efficient data structures. The data structure perspective to the performance of Spark-based Apriori has not been explored well. Singh et al. [19] have evaluated the performance of the Apriori algorithm on the different data structures, but on the Hadoop MapReduce and not on the Spark.

The original Apriori algorithm proposed by Agrawal and Srikant [1] uses Hash Tree data structure for the support counting, candidate generation and storage. Bodon and Rónyai [20] proposed an alternative data structure, Trie (Prefix Tree) for the same, which performs better that hash tree. Further, Bodon [21] has experimented hashing technique on the Trie (we named it as a Hash Table Trie), theoretically it accelerates the performance, but experimentally failed to perform, due to some memory constraints. Singh et al. [19] has investigated the effect of these three data structures on the MapReduce-based Apriori, and reported that the performance of Hash Table Trie is outstanding, while that of hash tree is worst. This paper re-investigates the study made by Singh et al. [19], but for the Spark-based Apriori, YAFIM [12]; and also discusses the order of variation of experimental results on the two big data processing platforms with respect to the data structures. Experimental results show that YAFIM with Trie and Hash Table Trie performs similar but many times better than Hash Tree for the Spark-based Apriori.

The major contributions of this paper are as follows:

1. Spark-based Apriori similar to YAFIM has been implemented using RDD framework, named as RDD-Apriori.

2. Two other variants of RDD-Apriori named as RDD-Trie-Apriori and RDD-HashTableTrie-Apriori have been developed by replacing the underlying data structure Hash Tree by Trie and Hash Table Trie respectively.

3. The performance of the three variants of RDD-Apriori is compared with each other on the both synthetic and real-life benchmark datasets.

4. The experimental results of these three variants of Spark-based Apriori have been analyzed and discussed with respect to the MapReduce-based Apriori in the context of data structures.

The rest of this paper is organized as follows. Section 2 is a preliminary section that describes frequent itemset mining and Apriori algorithm, and Apache Spark. Section 3 briefly discusses the related work. Implementation details of RDD-Apriori and description of Hash Tree, Trie, and Hash Table Trie data structures are given in section 4. Section 5 presents the experimental results and analysis. Finally, paper is concluded in section 6.

## 2. Preliminaries

### 2.1 Frequent Itemset Mining and Apriori Algorithm

Let $I = \{i_1, i_2, ..., i_m\}$ be a set of $m$ distinct attributes or items. A k-itemset is a set of $k$ items from the set $I$. If all items of an itemset $X$ are present in the itemset $Y$, then $X$ is called subset of $Y$, denoted as $X \subseteq Y$. A database $D$ is a set of $n$ transactions, denoted as $D = \{T_1, T_2, ..., T_n\}$, where each transaction $T_i$ consist of an itemset along with an unique transaction identifier $TID_i$ i.e. in the form of $<TID_i, i_1, i_2, ..., i_k>$. A transaction $T_i$ contains an itemset $X$ if $X \subseteq T_i$. The support count of $X$ denoted as $\sigma(X)$ is the occurrence frequency of $X$ i.e. the number of transactions containing $X$. If $\sigma(X) \geq min\_sup$, where $min\_sup$ is a user specified threshold value called minimum support, then $X$ is called frequent itemset [1]. Association rule mining [1] is a two step process. The first step generates all frequent itemsets, and the second step simply produces all the strong association rules from frequent itemsets. An association rule is a conditional implication of the form $X => Y$, where $X, Y \subset I$, and $X \cap Y = \phi$. The confidence of the rule $X => Y$ is a conditional probability, defined as $\sigma(X \cup Y) / \sigma(X)$. If confidence of the rule is more than or equal to $min\_conf$, where $min\_conf$ is a user specified threshold value called minimum confidence, the it is called strong rule. The generation of all frequent itemsets is a computation and memory intensive task. Apriori [1], Eclat [2], and FP-Growth [3] are the representative algorithms for frequent itemsets generation. Apriori is the simplest algorithm and easy to parallelize.

Apriori algorithm is founded on the Apriori property. Apriori property reduces the computation cost of support counting and the number of candidate itemsets. The candidate itemsets having minimum support are called frequent itemsets. The algorithm starts with generating frequent items $L_1$ by scanning the entire database. Now, for each $k^{th}$ iteration (k ≥ 2), frequent (k-1)-itemsets $L_{k-1}$ of previous iteration are joined conditionally with itself to generate candidate k-itemsets $C_k$; then Apriori property is applied to reduce the number of candidates. As per this property all the candidates will be eliminated from $C_k$ if any of its (k-1)-subset does not belong to $L_{k-1}$. The entire database is scanned for the checking of subset of each transaction against candidates to get the frequent k-itemsets [1]. An efficient and compact data structure plays a vital role in order to reduce such computation cost and memory requirement in the Apriori algorithm.

## 2.2 Apache Spark

Spark was originally developed at the AMPLab of University of California, Berkeley [10-11], to overcome the limitations of Hadoop MapReduce [5, 7] and to provide extremely faster and general cluster computing platform for big data processing. It was donated to the Apache Software Foundation in 2013 [9]. It does not only support batch processing of big data, but also an excellent platform for streaming data processing, iterative and interactive analytics. It provides scalability and fault tolerance similar to the Hadoop. The fundamental design principle of Spark is the Resilient Distributed Datasets (RDDs) [11]. Basically, RDD is a distributed memory abstraction as a collection of immutable data objects partitioned across the cluster nodes. Spark maintains a lineage chain that helps in fault tolerance. A lineage chain keeps the record of how an RDD is derived from another RDD i.e. the set of dependencies on the parent RDDs. If a partition of RDD is lost, then it can be rebuilt quickly through the lineage chain [11].

A spark cluster has a master node and a number of worker nodes, and they need not to be on different machines, for example, all can be on the same workstation [22]. The driver program of a Spark application controls the execution flow and the different operations in parallel on the Spark cluster. An application runs as a multiple parallel tasks. A task is a basic unit of work. A driver program launches and manages a number of executors on the worker nodes for an application. Executors are the processes running the tasks and keeping the data in memory or disk [22].

There are four ways to create an RDD. The first two are by parallelizing a collection, and from an external storage e.g. HDFS. The other two ways create RDD from another RDD, for example, by the transformation of an existing RDD, and by changing the persistence of an existing RDD [10]. The operations on RDDs are of two types: transformations and actions. The transformation operations create a new RDD by transforming an existing one. For example, some transformations are: map, flatMap, filter, reduceByKey. The action operations write the results to the external storage or return to the driver program, after applying the function of action. For example, some action operations are: collect, count, saveAsTextFile. Spark follows the lazy evaluation technique on the chain of transformations and action. Until the action is not triggered, no transformations will execute [23]. To provide communication among workers, Spark supports shared variables of two types: read only and write only. The read only shared variables are called broadcast variables and write only are called accumulators [10].

## 3. Related Work

Apriori [1], Eclat [2], and FP-Growth [3] are the central algorithms for frequent itemset mining. Researchers have applied more efficient techniques and data structures, and developed a number of improved versions of these basic sequential algorithms. Parallel and distributed algorithms have been developed to overcome the bottleneck of sequential algorithms [4]. As the more efficient and scalable distributed computing systems have been evolved with the time, designing of FIM algorithms has been shifted on those systems. After the emergence of Hadoop, a number of FIM algorithms have been redesigned on the Hadoop MapReduce framework. Due to being easy to parallelize, Apriori-based algorithms on MapReduce framework are explored more number of times than Eclat and FP-Growth based. Some well recognized and appealing MapReduce-based Apriori algorithms are PARMA, a parallel randomized algorithm on the MapReduce framework proposed by Riondato et al. [24], FPC (Fixed Passes Combined-counting), and DPC (Dynamic Passes Combined-counting) proposed by Lin et al. [25] and their improved and optimized versions by Singh et al. [26], and BIGMiner proposed by Chon and Kim [27]. The effect of data structures on MapReduce-based Apriori have been investigated by Singh et al. [19].

Dist-Eclat and BigFIM are the two Eclat-based algorithms proposed by Moens et al. [28]. Dist-Eclat is completely based on the Eclat while BigFIM uses Apriori approach initially and then Eclat. Peclat (Parallel Eclat) proposed by Liu et al. [29] is another Eclat-based algorithm on MapReduce framework that uses mixset approaches. A mixset dynamically select tidset or diffset for an itemset. PFP (Parallel FP-Growth) [30] is the parallel FP-Growth algorithm on MapReduce framework. BPFP is a balanced PFP algorithm proposed by Zhou et al. [31]. FiDoop and its extension FiDoop-HD proposed by Xun et al. [32] uses an FIU-Tree (frequent items ultrametric tree) instead of FP-Tree.

Although, Hadoop is scalable and fault tolerant distributed system, but does not perform well on the iterative algorithms. For each of the iterations, a new MapReduce job is launched each time that slows down the overall execution, as well as the costly write/read operations on the disk for intermediate results. Spark has been developed to overcome all such kinds of problems. Spark provides in-memory computation, and is highly suitable for iterative algorithms. YAFIM (Yet Another Frequent Itemset Mining) was the first Spark-based Apriori algorithm proposed by Qiu et al. [12]. It performed about 25 times faster than MapReduce-based Apriori. It generates frequent items in the first phase and frequent (k+1)-itemsets, k ≥ 2 in the second phase. Rathee et al. [13] have proposed R-Apriori (Reduced-Apriori) that improves YAFIM by introducing an additional phase to reduce the computation of 2-itemsets generation. It uses a bloom filter instead of the hash tree. Adaptive-Miner [14] improves R-Apriori by making a dynamic selection between the conventional or reduced approach of R-Apriori for the candidate generation. Yang et al. [15] used the array vectors data structure to map the database into memory and then parallelized this improved Apriori on the Spark. DFIMA (Distributed Frequent Itemset Mining Algorithm) [16] is a Spark-based Apriori algorithm that uses a Boolean vector for the frequent items and a matrix-based pruning method to reduce the size of candidates. HFIM (Hybrid Frequent Itemset Mining) [17] is also an Apriori-based algorithm along with vertical format of the dataset that reduces the scanning of the dataset. It uses both horizontal and vertical dataset obtained by eliminating infrequent items, where horizontal dataset is distributed across the worker nodes and vertical dataset is shared. DFPS (Distributed FP-Growth Algorithm based on Spark) [18] starts with production of frequent items, then repartition the conditional pattern base, and then computes frequent itemsets from the individual partitions in parallel.

## 4. Implementation Details

RDD-Apriori is a short name given to the Apriori algorithm on the Spark RDD framework. We do not propose any new algorithm here, but do the analysis of execution time of RDD-Apriori for the three different underlying data structures named Hash Tree, Trie, and Hash Table Trie. Three variants of RDD-Apriori have been implemented named as RDD-HashTree-Apriori (similar to YAFIM), RDD-Trie-Apriori, and RDD-HashTableTrie-Apriori corresponding to the three data structures. YAFIM [12], an RDD-Apriori algorithm with hash tree data structure, is considered as a benchmark for this case study. Section 4.1 and 4.2 describe the RDD-Apriori, and associated data structures respectively. Algorithm uses various transformation and action operations on RDDs which is self-explainatory, and more details can be found in the documentation [23].

### 4.1 RDD-Apriori

The algorithm comprises of two phases: Phase-1 and Phase-2, described as pseudo codes in Algorithm 1 and Algorithm 2 respectively. Phase-1 produces frequent items only and Phase-2 generates frequent k-itemsets for k ≥ 2. In Phase-1, transactions of database are partitioned as RDD across the available cores in the cluster. The first transformation, *flatMap()* splits each transaction and returns RDD of items. The transformation *mapToPair()* produces *(item, 1)* pairs for each item, then *reduceByKey()* makes sum of all values with the same key. Finally, the *filter()* transformation removes those pairs whose values are less than the *min_sup*, and returns only pairs of frequent items and their support count. The action operation, *collect()* returns the frequent items do the driver for subsequent use in the next phase. The frequent items and their count can be also saved to the disk using the action *saveAsTextFile()*. The number of partitions which determine the level of parallelism is preserved through all transformations and actions. The corresponding lineage graph for RDDs in Phase-1 is shown in Fig. 1.

**Algorithm 1:** Phase-1 of RDD-Apriori
1:   RDD transactions = sc.textFile("database");
2:   RDD items = transactions.flatMap(t -> List(t.split(" ")));
3:   PairRDD itemPairs = items.mapToPair(item -> (item, 1));
4:   PairRDD itemCounts = itemPairs.reduceByKey(($v_1$, $v_2$) -> $v_1$ + $v_2$);
5:   PairRDD freqItemCounts = itemCounts.filter(itemCount -> itemCount._2 >= min_sup);
6:   freqItemCounts.saveAsTextFile("$L_1$");
7:   freqItemsetList = freqItemCounts.keys().collect();

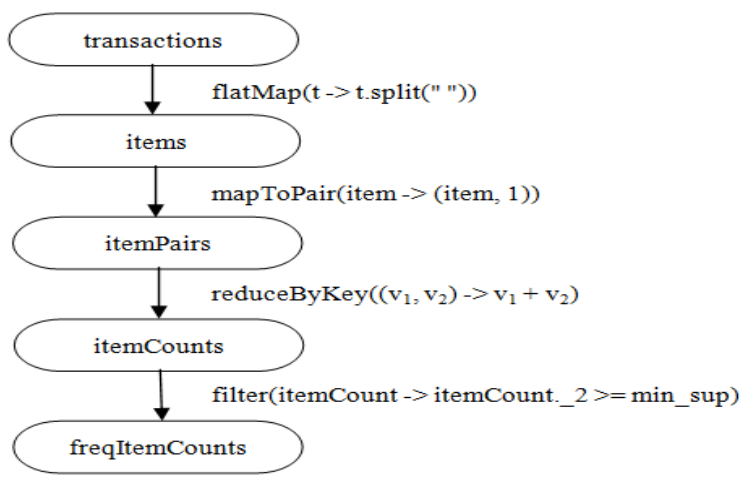

**Fig. 1:** Lineage graph for RDDs in Phase-1 of RDD-Apriori

Phaes-2 also applies the same sequence of transformations as in Phase-1 but iteratively. In the $k^{th}$ iteration, the frequent (k-1)-itemsets generated in the previous iteration are used to generate candidate k-itemsets. The frequent (k-1)-itemsets and candidate k-itemsets are stored in $treeL_{k\_1}$ and $treeC_k$ which may be a Hash Tree, Trie (prefix tree) or Hash Table Trie depending on the data structure used. The methods *apriori-gen()* and *subset()* are computationally intensive, so an efficient data structure should be used to significantly increase the performance. The *apriori-gen()* method generates the candidates by conditionally joining the frequent itemsets of previous iteration while *subset()* method checks the subset of each transaction against the candidates. The candidates stored in $treeC_k$ must be shared across all workers in order to apply *subset()* operation parallelly on the all partitions of transactions. The broadcast variable of Spark sends the copy of $treeC_k$ to all workers. The *flatMap()* transformation returns the list of candidate k-itemsets resulting from the *subset()* operation. The transformation *mapToPair()* converts each itemset into *(itemset, 1)* pairs, and the *reduceByKey()* reduces these pairs by summing up all values having the same key. Finally, the *filter()* operation eliminates the infrequent itemsets and returns only frequent k-itemsets. These frequent k-itemsets are saved to the disk and returned to the driver for subsequent use in the next iteration. Fig. 2 shows the lineage graph of RDDs of Phase-2. The source code of *apriori-gen()* and *subset()* methods based on hash tree have been taken from the SPMF Open-Source Data Mining Library [33].

**Algorithm 2:** Phase-2 of RDD-Apriori

```
1:  for(k = 2; !freqItemsetList.isEmpty; k++)
2:  {
3:      insert all itemsets of freqItemsetList into treeL_k_1
        // treeL_k_1 and treeC_k may be a hash tree, trie or hash table trie implemented separately
4:      treeC_k = apriori-gen(treeL_k_1);
5:      broadcast treeC_k to all workers
6:      RDD itemsets = transactions.flatMap(t -> {
7:          itemList = List(t.split(" "));
8:          if(itemList.size() >= k)
9:              cList = subset(treeC_k, itemList);
10:         return cList;
11:     });
12:     PairRDD itemsetPairs = itemsets.mapToPair(itemset -> (itemset, 1));
13:     PairRDD itemsetCounts = itemsetPairs.reduceByKey((v_1, v_2) -> v_1 + v_2);
14:     PairRDD freqItemsetCounts = itemsetCounts.filter(itemsetCount -> itemsetCount._2 >= min_sup);
15:     freqItemsetCounts.saveAsTextFile("/L"+_k);
16:     freqItemsetList = freqItemsetCounts.keys().collect();
17: }
```

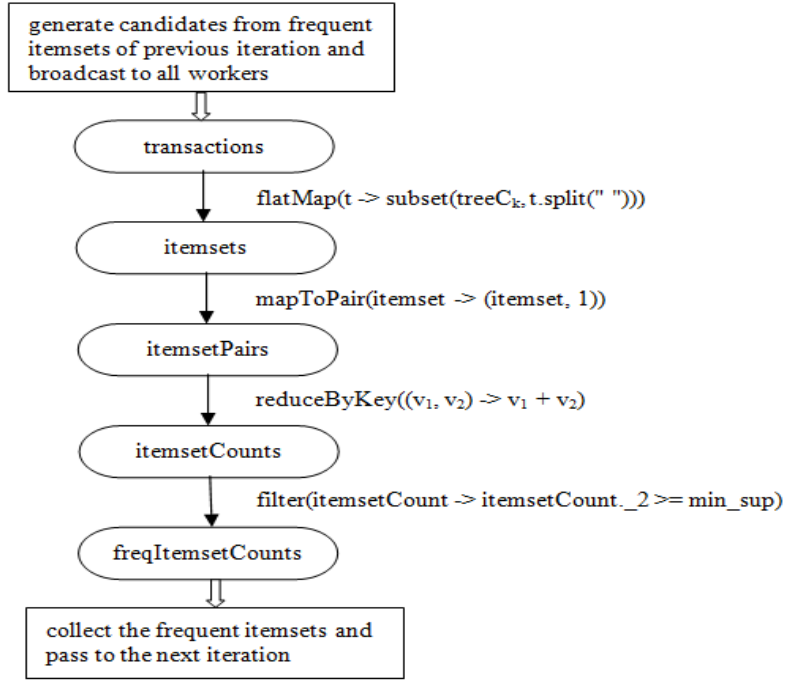

**Fig. 2:** Lineage graph for RDDs in Phase-2 of RDD-Apriori

### 4.2 Data Structures: Hash Tree, Trie, and Hash Table Trie

All the three data structures are tree with different properties. The common property is that traversing starts from the root node and goes downward towards the leaf nodes. The types of nodes in Hash Tree and Trie are different. There are two types of nodes in a Hash Tree: inner nodes and leaf nodes. The leaf nodes store the candidate itemsets while inner nodes directs to the child nodes using a hash function. Hash Tree also uses two parameter *child_max_size* and *leaf_max_size* that need to be set heuristically for the better performance. The parameter *child_max_size* is the maximum number of child nodes also called the table size for hashing. Whereas, the parameter *leaf_max_size* is a threshold value that controls the number of candidate itemsets at leaf nodes. A leaf node will be converted into an inner node if the number of candidates exceeds the value of *leaf_max_size* [20]. A prefix tree also known as Trie [20] which is a short name for the retrieval tree, stores the common prefixes only once. It does not store itemsets on nodes, but on the path from the root node to other nodes depending on the size of itemsets. An item is associated with each edge between the nodes at two consecutive levels. The use of prefix tree reduces the memory requirement, and the computation cost in candidate generation and support counting. Further, retrieving a candidate in a Trie is faster than Hash Tree. Hash Tree involves two steps, first hashing and then searching at the leaf node. Also, ist is easy and faster to generate candidates in a Trie on the basis of two common prefixes [20]. Fig. 3 (a) and (b) depict the Hash Tree and Trie respectively containing the same set of ten candidate 3-itemsets generated from the set of items, *{1, 2, 3, 4, 5}*. The set of candidate 3-itemsets is *{1 2 3; 1 2 4; 1 2 5; 1 3 4; 1 3 5; 1 4 5; 2 3 4; 2 3 5; 2 4 5; 3 4 5}*. The Trie in Fig. 3(b) stores these candidates on the basis of common prefixes. The Hash Tree in Fig. 3(a) uses a hash function defined over the items as *h(item) = item % child_max_size*, where *child_max_size* is 3. The source code of the Hash Tree data structure has been taken from the SPMF Open-Source Data Mining Library [33].

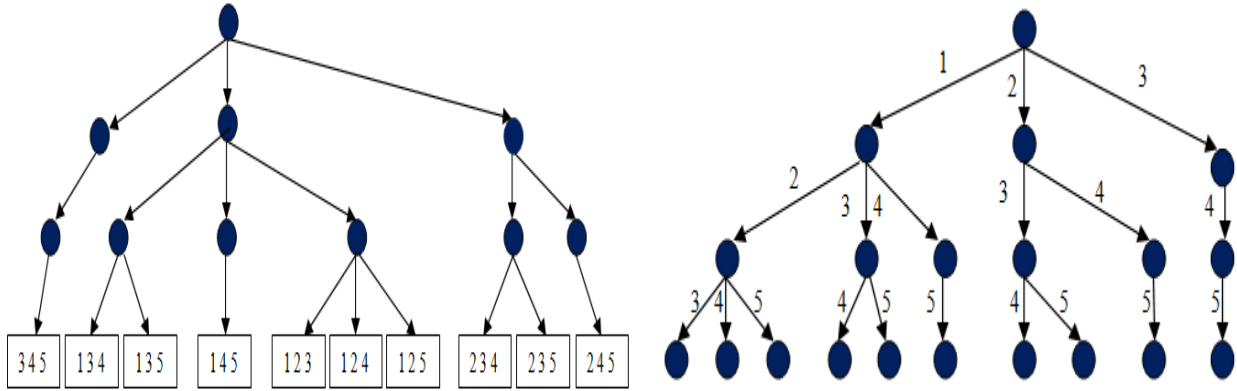

**Fig. 3:** (a) Hash Tree (b) Trie containing the same set of candidate 3-itemsets

Trie always does a linear search when moving downward from a node to any of the child nodes at the immediate lower level. Linear search at each step may increase the time of support counting. Hashing, an efficient searching technique may be employed to accelerate the search time. It is achieved by maintaining a hash table at each node in the Trie. A Trie with such hashing technique is named as a Hash Table Trie. Since each leaf node represents exactly one itemsets, so there must be a perfect hashing. In a Hash Table Trie, traversing from the root node to a leaf node requires a very small number of steps just equal to the width of itemset, which is possible only due to the hashing [21]. The only difference between Trie and Hash Table Trie is searching technique used for traversing downward, and the rest of all operations are the same in both data structures.

## 5. Experimental Results

### 5.1 Experimental Setup and Datasets

The setup is a local Spark cluster on a workstation having Intel Xenon CPU E5-2620@2.10 GHz with 24 cores, 16 GB RAM, and 1 TB of disk storage. Spark-2.1.1, Hadoop-2.6.0, Scala-2.11.8 are installed over Ubuntu 14.04 64 bit running on the workstation. Input datasets and generated frequent itemsets are stored in HDFS. All the source codes are written in Java-7. Algorithms are tested on a single workstation, but are scalable to the multi-nodes cluster.

Table 1 summarizes the seven benchmark datasets on which algorithms are evaluated. Synthetic as well as real-life, both kinds of datasets are used. The source of the datasets c20d10k, BMS_WebView_1, and BMS_WebView_2 (in short BMS1 and BMS2) is SPMF datasets [33] while that of the datasets chess, mushroom, T10I4D100K, and T40I10D100K is FIMI dataset repository [34].

**Table 1:** Datasets used in the experiments with their properties

| Dataset | Type of dataset | Transactions | Items | Average Transaction Width |
|---|---|---|---|---|
| c20d10k | Synthetic | 10,000 | 192 | 20 |
| chess | Real-life | 3196 | 75 | 37 |
| mushroom | Real-life | 8124 | 119 | 23 |
| BMS_WebView_1 | Real-life | 59602 | 497 | 2.5 |
| BMS_WebView_2 | Real-life | 77512 | 3340 | 5 |
| T10I4D100K | Synthetic | 100,000 | 870 | 10 |
| T40I10D100K | Synthetic | 100,000 | 1000 | 40 |

## 5.2 Performance Analysis

Performance of the three algorithms, RDD-HashTree-Apriori (existing YAFIM), RDD-Trie-Apriori (YAFIM with Trie data structure), and RDD-HashTableTrie-Apriori (YAFIM with Hash Table Trie data structure) have been evaluated in terms of the execution time. Performance of RDD-Trie-Apriori and RDD-HashTableTrie-Apriori is compared against the existing algorithm YAFIM or RDD-HashTree-Apriori. Figs. 4 (a-g) shows the execution times of the algorithms on varying value of minimum support for the seven datasets summarized in Table 1. It can be seen that on all the datasets, both RDD-Trie-Apriori and RDD-HashTableTrie-Apriori significantly outperform RDD-HashTree-Apriori. On the datasets BMS_WebView_1 and BMS_WebView_2, Trie and Hash Table Trie based Apriori outperform Hash Tree based Apriori by at least eight times (Figs. 4(d) and 4(e)), and about six times on the dataset T40I10D100K (Fig. 4(g)), at the lowest value of minimum support.

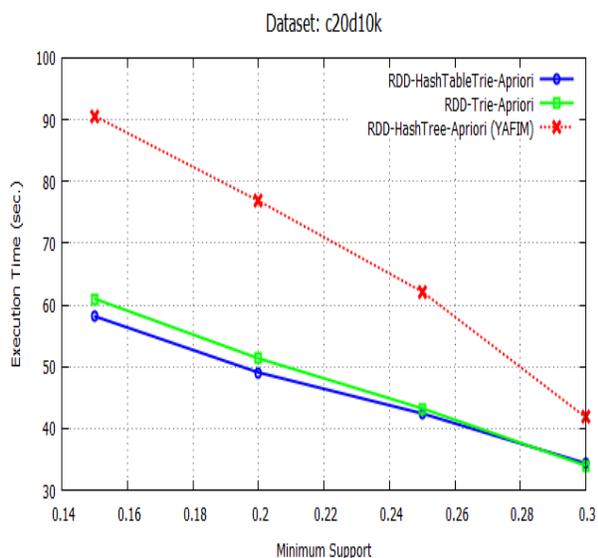

(a) Dataset c20d10k

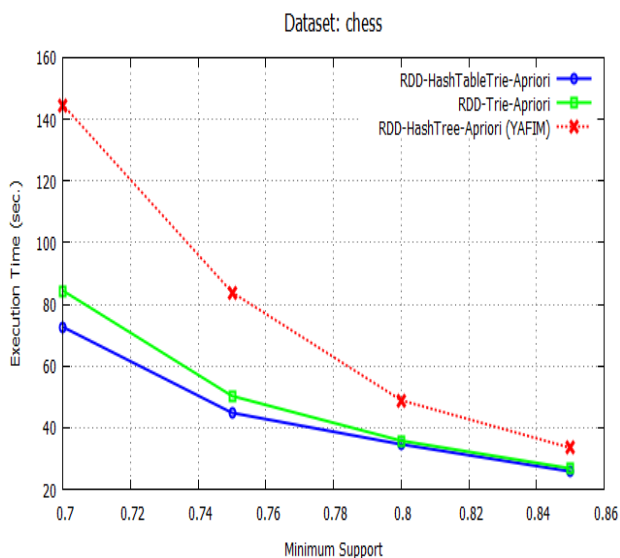

(b) Dataset chess

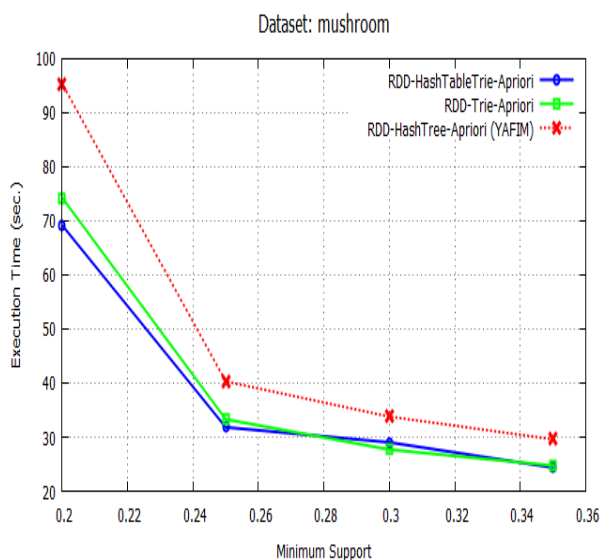

(c) Dataset mushroom

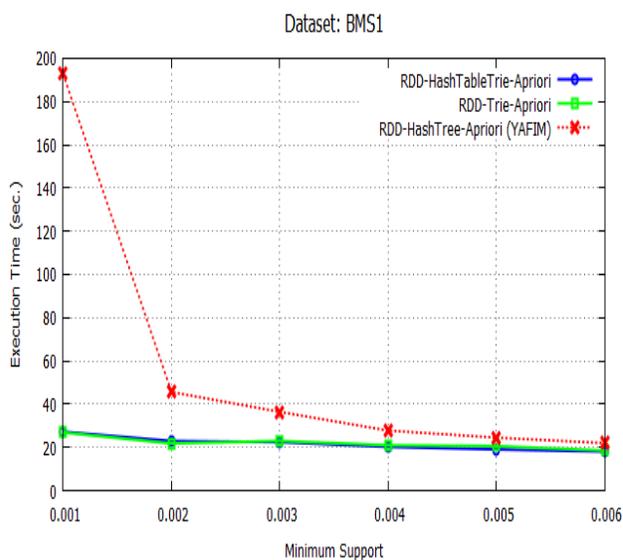

(d) Dataset BMS_WebView_1

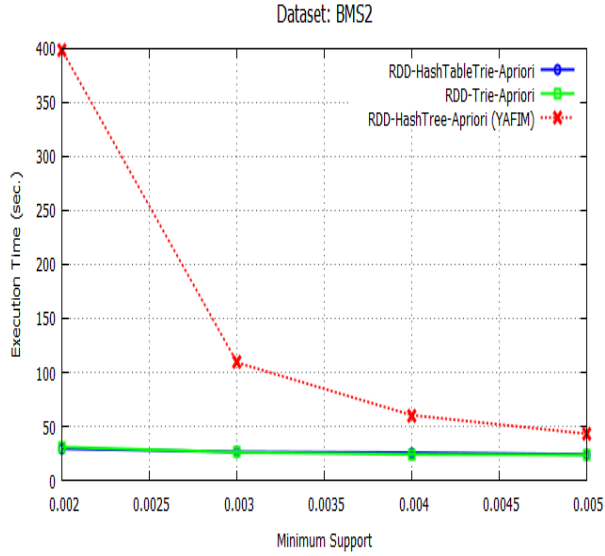

(e) Dataset BMS_WebView_2

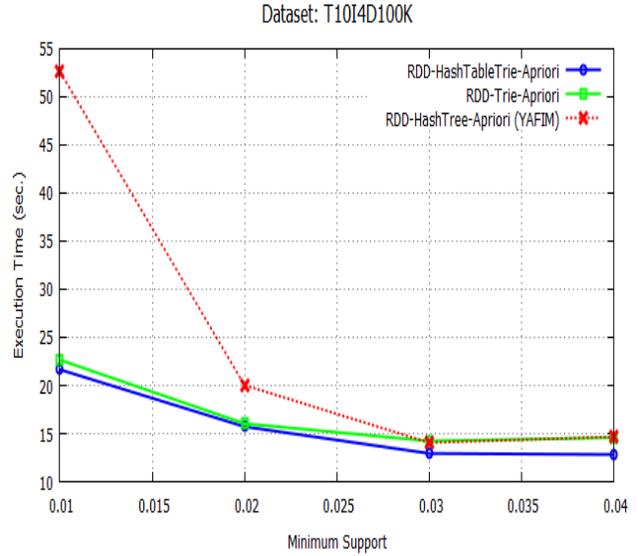

(f) Dataset T10I4D100K

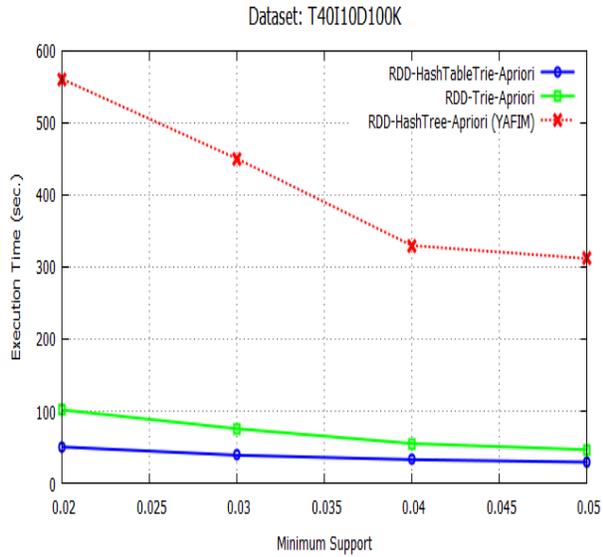

(g) Dataset T40I10D100K

**Fig. 4:** Execution time of algorithms RDD-HashTree-Apriori (YAFIM), RDD-Trie-Apriori, and RDD-HashTableTrie-Apriori for varying minimum support on seven datasets (a-g).

The difference between the execution time of RDD-Trie-Apriori and RDD-HashTableTrie-Apriori is not significant, even on some datasets their performance are similar to one another (Figs. 4(a-g)), but the both perform much better than the RDD-HashTree-Apriori. In the case of sequential Apriori implementation by Bodon [21], Hash Table Trie is theoretically sound but could not perform experimentally. The reason given by Bodon [21] for the failure of Hash Table Trie, is the larger size of its node due to hash table. A larger node could not be fit in the cache memory so reside in the main memory, and the read and write operations in the memory are slower than cache memory. The nodes of the Trie are simple which may be cached in, and the linear search is faster in cache memory. Singh et al. [19] have investigated the performance of the same three data structures in the context of MapReduce-based Apriori on the Hadoop cluster. The authors have shown in their experimental results that Hash Table Trie performed much better than Trie on some datasets while Hash Tree was the worst one. The most important reason behind the outstanding performance of Hash Table Trie on Hadoop MapReduce speculated by the authors, is the memory requirement during execution. Since, MapReduce launches a new job each time when a next iteration of Apriori begins. So, during the $k^{th}$ iteration, only the frequent (k-1)-itemsets of previous iteration and generated candidate k-itemsets are there to be residing in the memory; and for such small number of nodes, searching and reading operations become faster.

In the case of Spark-based Apriori, Hash Table Trie and Trie performs similarly, except on few datasets where Hash Table Trie is slightly better than Trie. Unlike MapReduce, Spark does not launch a new job each time when a next iteration of Apriori starts, but the transformation and action operations are repeated with the new iteration. The advantage of dedicated memory for the $k^{th}$ iteration is not possible in the Spark because it keeps all the data and the results of the previous iteration for performance optimization. Further, Spark takes the advantage of its distributed memory and RDD, so the performance of Hash Table Trie did not degrade like as in sequential Apriori, and perform equivalently to Trie.

## 6. Conclusion

This paper considered the designing of efficient Spark-based frequent itemset mining algorithm and in particular Spark-based Apriori. Many Spark-based Apriori algorithms have been proposed with the different approaches to parallelize the Apriori on Spark. Besides the parallelization strategies, the performance of the Apriori algorithm also depends on the efficient data structures like Hash Tree, Trie, and Hash Table Trie. Trie has been found as the most efficient data structure for the sequential implementation of Apriori. Hash Table Trie, a Trie with hashing technique, theoretically accelerates the sequential Apriori but failed experimentally. The most of the Spark-based Apriori algorithms have been designed with the Hash Tree and have not tried the other data structures. So, this paper has investigated the effect of Hash Tree, Trie, and Hash Table Trie on the Spark-based Apriori, and analyzed their relative performance with respect to the execution time. The experimentation was carried out on the seven benchmark datasets of both types i.e. synthetic and real-life, which shows that the Hash Table Trie and Trie based algorithm performs many times better than the Hash Tree based algorithm. Further, the Hash Table Trie based algorithm performs similar to the Trie based algorithm on some datasets and slightly better on the other datasets. Overall, the Apriori with Hash Table Trie excellently performs in the distributed memory environment of the Spark while it was the worst one in case of sequential Apriori.